\documentclass[conference]{IEEEtran}
\IEEEoverridecommandlockouts

\usepackage{cite}
\usepackage{color}
\usepackage{setspace}
\usepackage{amsmath,amssymb,amsfonts}
\usepackage{multirow}
\usepackage{graphicx}
\usepackage{textcomp}
\usepackage{xcolor}
\usepackage{hyperref}

\usepackage[linesnumbered, ruled]{algorithm2e}
\usepackage{setspace}
\usepackage{algpseudocode}
\algnewcommand{\LeftComment}[1]{\Statex \(\triangleright\) #1}

\usepackage{amsmath}

\def\BibTeX{{\rm B\kern-.05em{\sc i\kern-.025em b}\kern-.08em
    T\kern-.1667em\lower.7ex\hbox{E}\kern-.125emX}}
\begin{document}
\title{Agglomerative Federated Learning: \\ Empowering Larger Model Training via End-Edge-Cloud Collaboration
 %A Novel Framework for End-Edge-Cloud Collaborative Model Training
% \thanks{This work was supported by the National Key Research and Development Program of China (2021YFB2900102) and the National Natural Science Foundation of China (No. 62072436, No. 62002346 and No. 61872028).}
}
%Hierarchical Federated Learning for Large Model Training over Compute First Networking
\author{
	Zhiyuan Wu$^{1,2}$, Sheng Sun$^{1}$, Yuwei Wang$^1$\thanks{This paper has been accepted by IEEE International Conference on Computer Communications. \textit{(Corresponding author: Yuwei Wang)}}, Min Liu$^{1,3}$, Bo Gao$^{4}$, Quyang Pan$^{1,2}$, 
	\\ Tianliu He$^{1,2}$, and Xuefeng  Jiang$^{1,2}$\\
	$^1$Institute of Computing Technology, Chinese Academy of Sciences \\
	$^2$University of Chinese Academy of Sciences \quad $^3$Zhongguancun Laboratory \\
	$^4$Beijing Jiaotong University 
	 \\
	\{wuzhiyuan22s,sunsheng,ywwang,liumin\}@ict.ac.cn \quad	bogao@bjtu.edu.cn  \quad lightinshadow1110@gmail.com \\
	tianliu.he@foxmail.com \quad 
	jiangxuefeng21b@ict.ac.cn\\
}

\maketitle

%However, data sharing among devices may pose privacy challenges. To address this issue, 
\begin{abstract}
Federated Learning (FL) enables training Artificial Intelligence (AI) models over end devices without compromising their privacy. As computing tasks are increasingly performed by a combination of cloud, edge, and end devices, FL can benefit from this End-Edge-Cloud Collaboration (EECC) paradigm to achieve collaborative device-scale expansion with real-time access. Although Hierarchical Federated Learning (HFL) supports multi-tier model aggregation suitable for EECC, prior works assume the same model structure on all computing nodes, constraining the model scale by the weakest end devices. To address this issue, we propose Agglomerative Federated Learning (FedAgg), which is a novel EECC-empowered FL framework that allows the trained models from end, edge, to cloud to grow larger in size and stronger in generalization ability. FedAgg recursively organizes computing nodes among all tiers based on Bridge Sample Based Online Distillation Protocol (BSBODP), which enables every pair of parent-child computing nodes to mutually transfer and distill knowledge extracted from generated bridge samples. This design enhances the performance by exploiting the potential of larger models, with privacy constraints of FL and flexibility requirements of EECC both satisfied.
Experiments under various settings demonstrate that FedAgg outperforms state-of-the-art methods by an average of 4.53\% accuracy gains and remarkable improvements in convergence rate. Our code is available at \url{https://github.com/wuzhiyuan2000/FedAgg}.
\end{abstract}


%to collaborating with larger scale devices and analyzing massive amounts of data generated on the end side in real-time.
%With the rapid evolution of wireless networks and Internet of Things (IoT) technologies, massive amounts of data are generated at the network termination along with the explosive growth of mobile devices.
%Meanwhile, the unprecedented development of Artificial Intelligence (AI) simulates the potential of data analysis and processing for providing multitudinous intelligent services \cite{yao2022edge}.
%This drives a revolution in the mainstream computing paradigm from centralized cloud computing to decentralized End-Edge-Cloud Collaboration (EECC) \cite{duan2022distributed}, bringing the superiority of fully exploiting pervasive devices and rapidly accessing large amounts of data for delivering real-time services.
%Accordingly, distributed AI has become a promising trend to gradually replace conventional centralized AI paradigm \cite{santos2021data,rong2021edge}.

%Among the various distributed AI techniques, Federated Learning (FL) \cite{kairouz2021advances} enables collaborative training AI models over decentralized devices without exchanging their raw data, guaranteeing satisfactory model performance while protecting data privacy.

\section{Introduction}
Federated Learning (FL) \cite{kairouz2021advances,yang2019federated} has emerged as a promising technique for various Artificial Intelligence (AI) applications \cite{nguyen2021federated,fliotsurvey2} since it enables collaborative training of AI models over end devices without exchanging their raw data, ensuring both high model performance and data privacy.
As the mainstream computing paradigm shifts from centralized cloud computing to decentralized End-Edge-Cloud Collaboration (EECC) \cite{duan2022distributed}, large-scale computing tasks are increasingly performed by a combination of centralized cloud servers, bridge edge servers, and a large number of end devices instead of relying solely on the cloud.
FL can benefit from this more distributed computing paradigm to fully exploit pervasive devices and rapidly access large amounts of data for delivering real-time services \cite{bao2022federated}. 
However, prevailing FL methods \cite{mcmahan2017communication,li2020federated,karimireddy2020scaffold} adopt a simple client-server architecture with two tiers (also called levels), where the server is either on the cloud or the edge. The server collaborates with end devices (also called clients) by iteratively aggregating model parameters uploaded from devices and dispatching the updated global model to all devices. These FL methods are not suitable for EECC paradigm, which requires multi-level collaboration among computing nodes.

Fortunately, Hierarchical Federated Learning (HFL) \cite{liu2020client} is proposed to support multi-tier model aggregation in FL, from underlying devices to the bridged edge servers and the cloud servers.
This manner brings the advantages of device-scale expansion with multi-level collaboration.
When implementing HFL with EECC, computing power heterogeneity across end, edge, and cloud nodes should be taken into consideration to fully unleash the potential of EECC. 
Specifically, cloud servers are equipped with powerful computing capabilities and storage resources, which outperform edge servers, and end devices possess relatively constrained resources \cite{tak2020federated}.
However, existing HFL methods \cite{liu2020client,wang2022accelerating,wang2021resource,feng2022mobility} require imposing the same model structure on all computing nodes for model aggregation, inevitably limiting the scale of trained models by the weakest end devices due to the bottleneck effect and encountering resource under-utilization over edge and cloud computing nodes. 
To take full advantage of the computation superiority of edge and cloud nodes over end devices, it is critical to deploy appropriate models on end, edge, and cloud nodes that are compatible with their computation capability and also achieve model-agnostic \cite{makd} collaborative training in EECC-empowered FL methods.
In this way, larger models can be trained on the edge and cloud nodes with stronger characterization and generalization ability to achieve performance improvement.
%larger models can be trained with relatively limited end-side and edge resources, so as to fully exploit the hidden patterns in dispersed data and improve the generalization ability of the trained models.

%it is critical to train larger models in the edge and cloud nodes than on the end side that match their computation capability, aiming to fully mine valid information in dispersed data and improve model XX capability using larger models.
%improve model generalization capability, and thus enhance system performance.
%This prospect is a major challenge for existing HFL methods relying on multi-level model aggregation.}

%, impossible to train larger models on edge and cloud servers with sufficient capacity to fully mine valid information in dispersed data.
%Undoubtedly, small-scale models and under-utilized resources \textcolor{red}{over edge and nodes} hinder unleashing the advantages of HFL.
%Therefore, it is critical to train larger models compatible with the computation capability of nodes via HFL empowered by EECC to achieve satisfactory performance.

To this end, we propose a novel FL framework suitable for EECC paradigm, named Agglomerative Federated Learning (FedAgg).
To be specific, FedAgg recursively organizes computing nodes to collaboratively train models that grow in size from end devices, edge servers to cloud servers via our customized Bridge Sample Based Online Distillation Protocol (BSBODP), which defines the interaction rules among every pair of parent-child computing nodes in an end-edge-cloud network with a tree-topology. During training, the knowledge acquired by the upper-level nodes (closer to the cloud) from the lower-level nodes (closer to devices) is recursively transferred to the cloud, tier by tier via BSBODP.
In this way, upper-level nodes can deploy larger models with higher capability and integrate the knowledge of lower-level nodes in an agglomerative manner.
Regarding BSBODP, model-agnostic collaborative training is achieved in any pair of parent-child computing nodes, where a pre-trained lightweight decoder is used to generate fake samples (called bridge samples) whose extracted logits are used for knowledge distillation among nodes.
%where one computing node generates fake samples (called bridge samples) with a pre-trained lightweight decoder and then performs knowledge distillation on logits extracted from bridge samples over the other node.
To our best knowledge, 
\textbf{agglomerative federated learning is the first framework empowered by end-edge-cloud collaboration paradigm that enables training larger models with ever-increasing capability tier by tier up to the cloud.}
%\textbf{agglomerative federated learning is the first framework empowered by end-edge-cloud collaboration that allows larger models to be trained on the cloud beyond the maximum capacity of the edge servers and end devices.} 
The proposed framework is able to train models that are adapted to the power of computing nodes and satisfy the privacy constraints of FL as well as the flexibility requirements of EECC.
%achieve superior performance compared with state-of-the-art methods.

The main contributions of this paper are summarized as follows:
\begin{itemize}
    \item 
    We propose a novel FL framework suitable for EECC paradigm, named Agglomerative Federated Learning (FedAgg), which recursively organizes computing nodes via a customized interaction protocol and enables models trained on the end, edge, and cloud nodes to grow larger in size and stronger in generalization ability.
    \item 
    We design the Bridge Sample Based Online Distillation Protocol (BSBODP) to achieve model-agnostic collaborative training among interacted nodes via online distillation over generated bridge samples.
    \item 
    We validate the effectiveness of FedAgg in an end-edge-cloud architecture over CIFAR-10 and CIFAR-100 datasets. Empirical results demonstrate that FedAgg achieves superior accuracy and significantly faster convergence compared with related state-of-the-art methods.
\end{itemize}

\section{Related Work}
\subsection{Hierarchical Federated Learning}
Hierarchical Federated Learning (HFL) is first proposed by Liu \cite{liu2020client}, where the established end-edge-cloud FL architecture combines the advantages of both massive data coverage of cloud computing and low communication latency of edge computing. 
Following Liu \cite{liu2020client}, a series of works \cite{wang2022accelerating,wang2021resource,feng2022mobility} are proposed to accelerate the training process or improve the system performance of HFL by building clients' cluster topology and performing hierarchical aggregation on top of local training, taking into account the computing and communication capability \cite{wang2021resource,wang2022accelerating} as well as data distribution and mobility \cite{feng2022mobility} of end devices.
In particular, Nguyen \cite{nguyen2022self} proposes a novel hierarchical distributed learning framework inspired by democratized learning \cite{nguyen2021distributed}, which extends conventional HFL through a self-organizing hierarchical structure based on agglomerative clustering, and enhances both local personalization and global generalization in distributed training.
However, all the above methods require participating computing nodes to adopt the same model structure, which limits the model scale to the capability of end devices and wastes the resources of edge and cloud nodes with substantial computing power.
%This manner also prevents the HFL system from training large models that require on-device powerful resource support.

\subsection{Training Larger Model in Federated Learning}
With sufficient training data, increasing model size is a common way of improving model accuracy and generalization ability in deep learning \cite{hu2021model}. 
However, prevailing FL approaches \cite{mcmahan2017communication,li2020federated,karimireddy2020scaffold} require homogeneous models to be adopted in both the powerful central server and the resource-limited end devices, which prevents the training of larger models beyond the maximum capability of end devices. To address this challenge, prior works \cite{alam2022fedrolex,nguyen2022feddct,cho2022heterogeneous,he2020group,cheng2021fedgems,wu2022exploring} propose two types of solutions to train larger models that surpass the resource limits of devices, one based on partial training and the other based on knowledge distillation.
In particular, partial training-based approaches \cite{alam2022fedrolex,nguyen2022feddct} divide the complete global model into multiple sub-models and then train these sub-models on multiple clients in parallel, thereby allowing the model on server obtained by collaborative training to exceed the size of the largest model on clients.
Knowledge distillation-based approaches \cite{cho2022heterogeneous,he2020group,cheng2021fedgems,wu2022exploring} leverage the model output of clients as a regularization term for the server-side model training, thereby achieving the transfer of integrated representation learned by clients on decentralized private data.
However, the above methods are designed for two-tier architectures and cannot support collaborative model training empowered by end-edge-cloud networks. 
Therefore, the applicability of the aforementioned works is highly restricted.
%Therefore, the scalability of the previous works is largely constrained by the capacity of the edge servers.

% most of the above approaches \cite{alam2022fedrolex,nguyen2022feddct,cho2022heterogeneous,he2020group} \textcolor{red}{cannot support dynamic migration of computing nodes when scaling to a three-layer network due to the restrictions of the model structures on clients and the server.}
% Furthermore, the literature \cite{cheng2021fedgems} requires training with the auxiliary of a public dataset, which is impractical in real scenarios.

% \begin{figure*}[t]
% 	\centering
% 	\includegraphics[width=1.0\textwidth]{motivation.pdf}
% 	\caption{Dynamic migration of computing nodes and model scale deployment for different network tiers in end-edge-cloud collaboration.}
% 	\label{motivation}
% \end{figure*}

\section{Preliminary}
%\subsection{Notations and Definations}
%In this subsection, we provide notations and definitions related to End-Edge-Cloud Collaboration (EECC) and Hierarchical Federated Learning (HFL) empowered by EECC.
\subsection{End-Edge-Cloud Collaboration}
  \label{eecc}
    End-Edge-Cloud Collaboration (EECC) is an emerging computing paradigm that coordinates multi-level heterogeneous computing nodes including central cloud servers, numerous edge servers, and massively distributed end devices, aiming to collaboratively work on large-scale computing tasks. In general, end devices are at the periphery of the network, generating data to perform data-relevant AI applications.
    Edge servers bridge to connect end devices and cloud servers, distributed at intermediate locations along the network. 
    Cloud servers are equipped with sufficient computation and storage resources, enabling the coordination of underlying end devices and edge servers.
    % EECC, as shown in Fig. \ref{motivation} (a), is an emerging computing paradigm that coordinates multi-level computing nodes across end, edge, and cloud that communicate and collaborate with each other via network links, aiming to work on large-scale computing tasks with heterogeneous computing resources among layers fully used.
    %In our setting, the central nodes $\mathcal{V}_1$ furthest from the user and data source are called the cloud, and the end nodes $\mathcal{V}_t$ closest to the user and data source are called the end. In addition, the edge refers to the edge side of cloud computing, located at the intermediate node between the cloud and the end devices. Moreover, the edge consists of nodes that are located outside the cloud and nearer to the end devices, distributed at intermediate locations between the cloud and the end users.

\begin{figure}[t]
	\centering
	\includegraphics[width=0.5\textwidth]{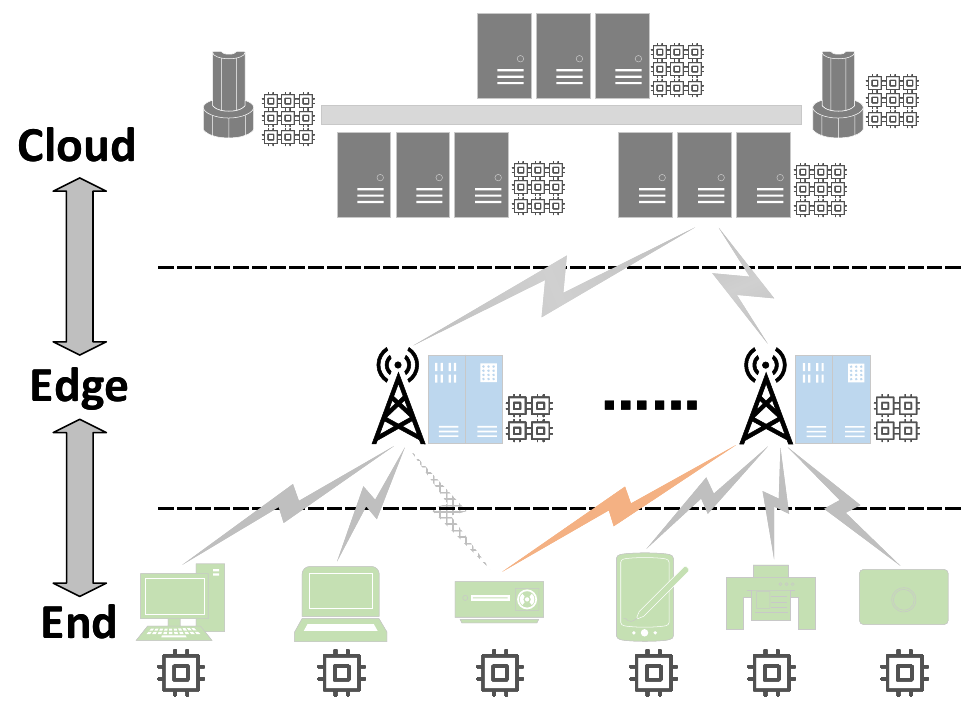}
	\caption{Framework of end-edge-cloud collaboration.}
	\label{framework}
\end{figure}

    We consider a typical End-Edge-Cloud Network (EEC-NET) with a tree topology (shown in Fig. \ref{framework}) denoted as $G=(\mathcal{V},\mathcal{E})$, where $\mathcal{V}$ is the set of computing nodes and $\mathcal{E}$ is the set of communication links.
    The entire EEC-NET $G$ has only one root node $r\in \mathcal{V}$ and contains one or more leaf nodes that form a set $\mathcal{L} \subset \mathcal{V}$. 
    Each computing node $v\in \mathcal{V}$ except leaf nodes has one or more children, and all of which form a set $Child(v)$. 
    Any computing node $v\in \mathcal{V}$ except the root node has one parent $Parent(v)$. 
    Besides, we define $Leaf(v)$ to be all the leaf nodes of the sub-tree that seeks the computing node $v$ as its root, and hence we have $Leaf(r)=\mathcal{L}$. 
    Without loss of generality, the computing nodes in EEC-NET are arranged by levels. Assuming a EEC-NET $G$ have a total of $T$ tiers, and the computing nodes in tier $t\in\{1,2,...,T\}$ form a set $\mathcal{V}_t$, so that we have $\mathcal{V}_T=\mathcal{L}$ and $\mathcal{V}_1=\{r\}$.
    For flexibility reasons, EECC should take into account the dynamic migration of computing nodes caused by manifold factors such as uneven load distribution, unreliable network connections, node failures, etc. Specifically, each node should be able to flexibly switch to different parent nodes at the same level, as shown in Fig. \ref{framework}.
    %Hence, it is practical to assume that an arbitrary computing node can switch between different parents of nodes at the same level, which requires distributed algorithms via EECC to allow for flexible switching of relationships between computing nodes.

    \subsection{Hierarchical Federated Learning Empowered by End-Edge-Cloud Collaboration}
    %FL is a distributed machine learning technique that collaboratively trains models on multiple computing nodes without sharing the raw data of participants.
    %We consider training a machine learning model via EECC, where there are $K=|L|$ end devices (clients) participating in FL and each client $k\in\{1,2,...,K\}$ owns a local dataset ${\mathcal{D}^k} = \bigcup\limits_{i = 1}^{{N^k}} {\{ (X_i^k,y_i^k)\} }$, in which $N^k$ is the number of samples in $\mathcal{D}^k$, $X_i^k$, $y_i^k$ is the data and label of the i-th sample in $\mathcal{D}^k$.
    Assume $K=\mid \mathcal{L}\mid$ end devices (called clients) participate in Hierarchical Federated Learning (HFL) over an EEC-NET, and each client $k\in\{1,2,...,K\}$ owns a local dataset ${\mathcal{D}^k} = \bigcup\limits_{i = 1}^{{N^k}} {\{ (X_i^k,y_i^k)\} }$, in which $N^k$ is the number of samples in $\mathcal{D}^k$, and $X_i^k$, $y_i^k$ are the data and label of the $i$-th sample in $\mathcal{D}^k$, respectively.
    Besides, each node $v\in \mathcal{V}$ holds a model $Model(v)$ with parameters $W^v$, and enables to exchange information with its parent and child nodes, i.e., $\{Parent(v)\} \cup Child(v)$.
    Our goal is to obtain the model in the root node $Model(r)$ that minimizes its training loss $L_{train}(\cdot)$ over all private data on devices, that is:
    %FL via EECC requires holding a model $Model(v)$ on each computing node $v\in \mathcal{V}$ with parameters noted as $W^v$. Each node $v\in \mathcal{V}$ can exchange information with the neighboring nodes $\{Parent(v)\} \cup Child(v)$. Our goal is to train out a model in the cloud $Model(r)$ that minimizes the loss over all private data on devices, that is:
	\begin{equation}
		\mathop {{\rm{argmin}}}\limits_{{W^r}} {L_{train}}(\bigcup\limits_{k = 1}^K {{\mathcal{D}^k}} ;{W^r}).
	\end{equation}

\begin{figure}[t]
	\centering
    \includegraphics[width=0.375\textwidth]{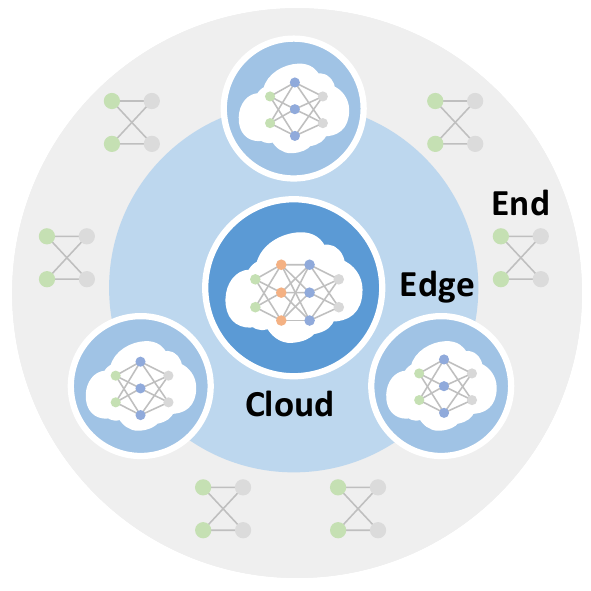}
	\caption{Expected model scales on different tiers in an end-edge-cloud network.}
	\label{model}
\end{figure}

    To fully exploit heterogeneous resources and accommodate the differentiated capabilities of end, edge, and cloud computing nodes, the edge and cloud servers with powerful computation resources should deploy larger models than end devices to enhance performance by capturing more complex and generalized patterns from private data, while the model size on end nodes with limited capability should be reduced accordingly, as shown in Fig. \ref{model}.

    %\textcolor{red}{Therefore, it is challenging to conduct large-scale model training in HFL under the constraints of end-edge-side small-scale models.} 
    %In particular, computing resources typically grow as they are distant from the data source, thus requiring larger models to be deployed away from the data to adapt to the power of computing nodes, as shown in Fig. \ref{motivation} (c).
    %end devices such as mobile phones, tablets, and smartwatches are commonly resource-limited and have high individual variability in hardware supports \cite{tak2020federated}. Meanwhile, the model in the cloud need to possess a scale well beyond the largest model on the end side to achieve better performance and generalization.
    %Therefore, this requires FL via EECC to allow for large-scale cloud models to be trained within the constraints of end devices.

\subsection{Interaction Protocols in Federated Learning}
Interaction protocols in FL are rules of information exchange between computing nodes in an FL system, including rules for both content transfer and parameters update. 
Prevailing FL interaction protocols are based on parameters interaction like FedAvg \cite{mcmahan2017communication}, where upper-level nodes aggregate and distribute model parameters from and to lower-level nodes.
However, this protocol requires all computing nodes to deploy a model with the same structure, which limits the model size on the cloud or the edge by the capacity of end devices.

\begin{figure*}[t]
	\centering
	\includegraphics[width=1.0\textwidth]{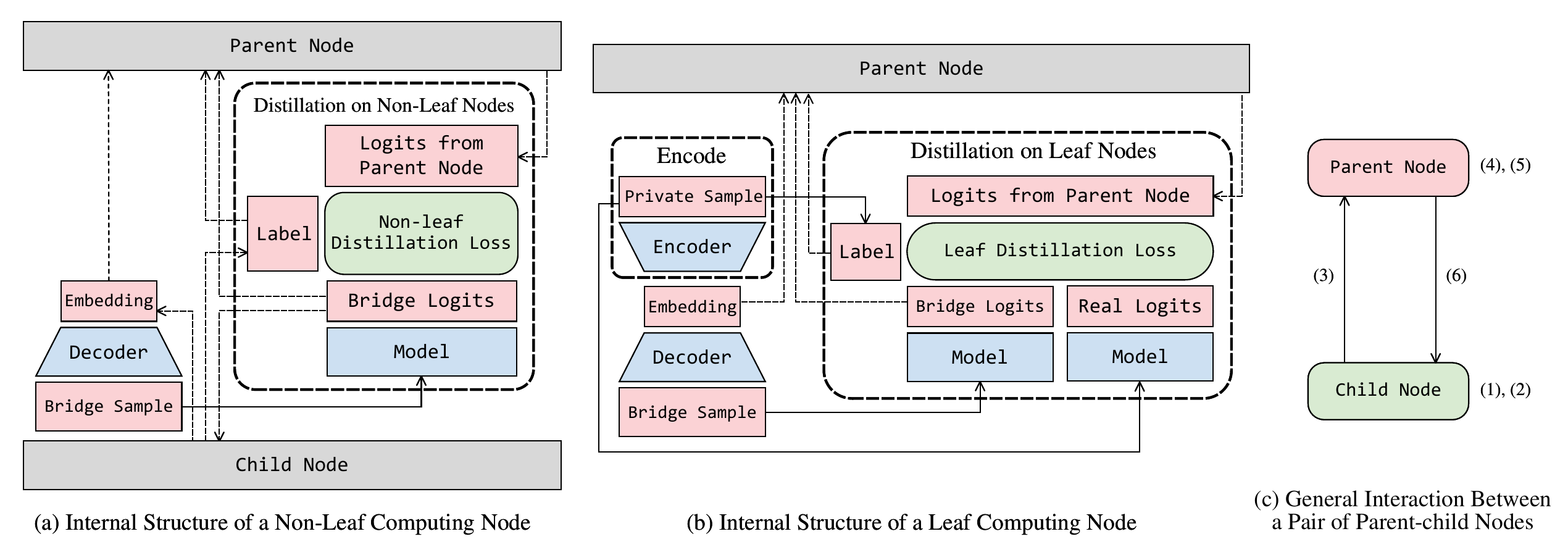}
	\caption{Overview of BSBODP. 
		(1) Child Node Distillation on Bridge Samples.
		(2) Logits Extraction on Child Node.
		(3) Upload Logits to the Parent Node.
		(4) Parent Node Distillation on Bridge Samples. 
		(5) Logits Extraction on Parent Node.
		(6) Distribute Logits to the Child Node.}
	\label{bsodp}
\end{figure*}

\begin{figure}[t]
	\centering
	\includegraphics[width=0.5\textwidth]{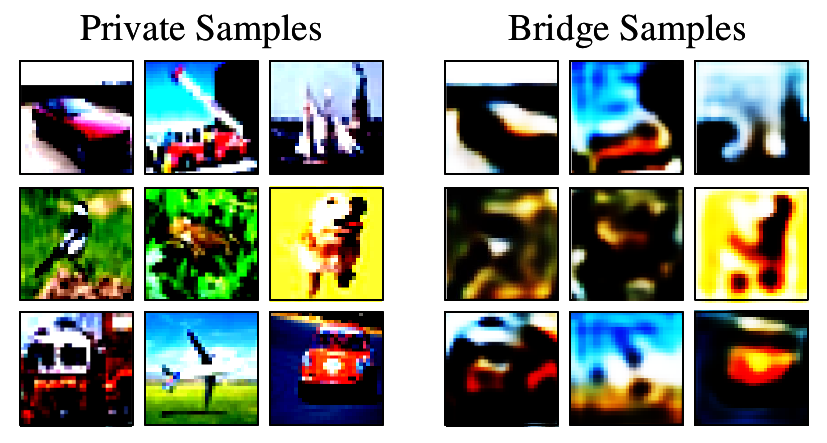}
	\caption{Comparison of private samples and bridge samples.}
	\label{bridge}
\end{figure}

\section{Agglomerative Federated Learning}
\subsection{Motivation and Overview}

Considering the intrinsic property of EECC, lower-level end devices possess abundant data but limited computing resources, while the upper-level nodes have powerful computation ability but no data.
In order to access local data directly and meet the privacy requirements of FL, existing EECC-empowered FL methods \cite{liu2020client,wang2022accelerating,wang2021resource,feng2022mobility} conduct model training on end devices and only perform model aggregation on the edge and cloud.
However, this manner requires the same model structure to be adopted among all computing nodes, thereby limiting the scale of trained models by the weakest end devices and encountering computation capability waste of edge and cloud nodes.

To exploit the powerful computation of edge and cloud nodes, we prompt the intuitive motivation of training larger models on edge and cloud nodes than end devices for achieving higher accuracy with stronger generalization abilities.
A feasible way to realize this idea is to conduct 
online distillation \cite{anil2018large,wang2021knowledge} as an interaction protocol among computing nodes at different tiers, which enables different nodes to train models with heterogeneous structures via iteratively exchanging the logits (also called knowledge) between two nodes to guide reciprocal model training. 
However, directly integrating online distillation into EECC-powered FL may raise privacy concerns about sharing data on devices, as the same sample is required to extract logits across different computing nodes.
%requires model training on two computing nodes over logits extracted from the same sample, raising privacy concerns as data is not being allowed to leave devices.
To overcome this, we design a new interaction protocol named Bridge Sample Based Online Distillation Protocol (BSBODP), which employs fake data as a bridge to transfer knowledge across multi-level nodes with the assistance of a lightweight pre-trained autoencoder.
%The logits extracted by fake data generated from the decoder from either encoded or received embeddings will not reveal raw data on devices.
This fake data is generated by the decoder from either encoded or received embeddings, allowing for computing logits without revealing raw data on devices.

Given a multi-tier architecture in EECC, each node can adopt a model structure that matches its own computing power, and the lower-level nodes can iteratively transfer their learned representations to the upper-level nodes via BSBODP. 
Therefore, the model on the cloud can eventually integrate the knowledge of all nodes and achieve satisfactory performance with sufficient capacity.
Based on this concept, we propose Agglomerative Federated Learning (FedAgg), which enables the training of larger models with better performance on powerful cloud and edge nodes than on resource-limited end devices via agglomerating knowledge generated by ever-expanding models from bottom to top in the multi-tier architecture in EECC. 
By leveraging the advantages of knowledge agglomeration, FedAgg enables the trained models from the bottom to the top to grow in size and generalization ability without violating the privacy principle in FL. Moreover, FedAgg can also handle scenarios where any computing nodes dynamically change their parents within the same tier, which ensures deployment flexibility in a realistic EEC-NET.

\subsection{Bridge Sample Based Online Distillation Protocol}
Fig. \ref{bsodp} illustrates the overview of our designed BSBODP, where each computing node preserves a pre-trained decoder and a conventional complete model (the model to be trained on the computing node).
In particular, each leaf computing node additionally preserves a pre-trained encoder (forms a pre-trained autoencoder with the aforementioned decoder) for encoding local data into embeddings, which will be transmitted to its parent node until reaching the root node.
During the execution phase, each computing node utilizes the decoder to generate bridge samples that match the data distribution of its corresponding leaf nodes based on the embeddings encoded from private data or uploaded by sub-nodes.
These bridge samples are used as intermediaries to conduct online distillation between every pair of parent-child computing nodes aligned by the same embedding.
As illustrated in Fig. \ref{bridge}, it is difficult to recover raw information from the embeddings of private samples since the embeddings are generated based on an extremely lightweight autoencoder ($<$50K model parameters) that are pre-trained on a super large open dataset (such as ImageNet \cite{deng2009imagenet}). As the autoencoder is not able to capture the fine-grained features and reconstruct training samples from them, bridge sample generation and distillation will not reveal data privacy.

\begin{algorithm}[t]
\newcommand{\removelatexerror}{\let\@latex@error\@gobble}
	\caption{Bridge Sample Based Online Distillation Protocol (BSBODP)}
	\SetAlgoNoLine
	\LinesNumbered

	\label{alg1}
	\begin{spacing}{0.72}
		\KwIn{$v^1$,$v^2$}%输入参数
		\KwOut{Trained $Model(v^1)$, $Model(v^2)$}%输出
        \begin{algorithmic}[1]
            \Procedure{BSBODP}{$v^1$, $v^2$}
		%\textbf{procedure BSBODP($v^1$, $v^2$)}:\\
		%\quad \textbf{BSBODP-Directional($v^1$, $v^2$)}\\
           \State \Call{BSBODP-Directional}{$v^1$, $v^2$}
		%\quad \textbf{BSBODP-Directional($v^2$, $v^1$)}\\
          \State \Call{BSBODP-Directional}{$v^2$, $v^1$}
          %\quad \textbf{return} $Model(v^1)$, $Model(v^2)$
          \State \Return{$Model(v^1)$, $Model(v^2)$}
		\EndProcedure

           \Procedure{BSBODP-Directional}{$v^{S}$, $v^{T}$}
            \State 
                $v^{T}$ generates bridge samples $dec(\varepsilon)$ from all stored embeddings $\varepsilon$
            \State $v^{T}$ extracts logits ${z^\varepsilon} = f(dec(\varepsilon );{W^{{v^T}}})$ on bridge samples, and transmits the results to $v^{S}$
            \State \If{$v^{S} \notin \mathcal{L}$}
            {
            \State Optimize $W^{v^{S}}$ according to Eq. (\ref{optim-non-leaf})
            }
            \State \Else
            {
            \State Optimize $W^{v^{S}}$ according to Eq. (\ref{optim-leaf})
            }
            \EndProcedure   
           \end{algorithmic}
	\end{spacing}
\end{algorithm}

\begin{algorithm}[!t]
	\caption{Agglomerative Federated Learning (FedAgg)}
	\SetAlgoNoLine
	\LinesNotNumbered
	
	\label{alg2}
	\begin{spacing}{0.72}
		\KwIn{$\mathcal{V},\mathcal{E}$}%输入参数
		\KwOut{Trained $Model(v^*), \forall v^*\in \mathcal{V}$}%输出
  \begin{algorithmic}[1]
      \Procedure{FedAgg}{$\mathcal{V},\mathcal{E}$}
      \State \Call{Init}{$r,\mathcal{E}$}
      \State \While{do not reach maximum epoches}
      {
      \State \Call{FedAggTrain}{$r,\mathcal{E}$}
      }
      \EndProcedure

      \Procedure{Init}{$v^*,\mathcal{E}$}
      \State \If{$v^*=r$}
      {
      \State \For{$u \in Child(v^*)$ in parallel}{
      \State \Call{Init}{$u,\mathcal{E}$}
      \State Receive and store embeddings $\varepsilon$ with corresponding labels from $u$}
      }
      \State \ElseIf{$v^* \in \mathcal{L}$}
      {
      \State Extract and store embeddings $\varepsilon$ via $enc(X^*),\forall X^*\in \mathcal{D}^{v^*}$
      
      \State Transmit embeddings $\varepsilon$ with corresponding labels to $Parent(v^*)$
      }
      \State \Else
      {
          \For{$u \in Child(v^*)$ in parallel}
          {
          \State \Call{Init}{$u,\mathcal{E}$}
          \State Receive and store embeddings $\varepsilon$ from $u$
          \State Send embeddings $\varepsilon$ with corresponding labels to $Parent(v^*)$
          }
      }
      \EndProcedure

      \Procedure{FedAggTrain}{$v^*,\mathcal{E}$}
      \State \If{$v^*=r$}
      {
      \State \For{$u \in Child(v^*)$ in parallel}
      {
      \State \Call{FedAggTrain}{$u,\mathcal{E}$}
      }
      }
      \State \ElseIf{$v^*\in \mathcal{L}$}
      {\Call{BSBODP}{$v^*$,$Parent(v^*)$}}
      \State \Else
      {
      \State \For{$u \in Child(v^*)$ in parallel}
      {
            \State \Call{FedAggTrain}{$u,\mathcal{E}$}
            \State \Call{BSBODP}{$v^*$,$Parent(v^*)$}
      }
      }
      \EndProcedure
      % \textcolor{red}{\Procedure{BSBODP}{$v^1,v^2$}
      %   \State Referencing to Algorithm \ref{alg1}, \Call{BSBODP}{$v^1,v^2$},
      % \EndProcedure}
  \end{algorithmic}	
%		\quad \textbf{BSODP-DIRECTIONAL($\mathcal{V}_1$, $v_2$)}\\
%		\quad \textbf{BSODP-DIRECTIONAL($v_2$, $\mathcal{V}_1$)}\\
%		\quad \textbf{return} $Model(\mathcal{V}_1)$, $Model(v_2)$
%		
%		
%		\textbf{procedure BSODP-DIRECTIONAL($v^{S}$, $v^{T}$):}\\
%		\quad $v^{T}$ extract logits $z^{\varepsilon}$ from $dec(\varepsilon)$\\
%		\quad $v^{T}$ send logits $z^{\varepsilon}$ to $v^{S}$\\
%		\quad \textbf{if }{$v^{S} \notin L$}\\
%		{
%			\quad \quad 	Optimize $W^{v^{S}}$ according to Eq. (\ref{optim-non-leaf})\\
%		}
%		\quad \textbf{else }\\
%		{
%			\quad \quad 	Optimize $W^{v^{S}}$ according to Eq. (\ref{optim-leaf})\\
%		}
%		\quad \textbf{end}
	\end{spacing}
\end{algorithm}
%%%%%%%%%%%%%%%%%%%%%%%%%%%%%%%%%%%%%%%%%%%%%%%

We formulate the process of BSBODP as follows: the decoder on each computing node is defined as $dec(\cdot)$, and the encoder on each leaf computing node is defined as $enc(\cdot)$. 
In addition, the model to be trained on any computing node $Model(v^*),v^* \in \mathcal{V}$ can give an inference on the input data via $f(\cdot;W^{v^*})$.
%Specifically, we define the decoder on each computing node as $dec(\cdot)$, and the model on any computing node $v^*$ can give an inference on the input data via $f(\cdot;W^{v^*})$.
When performing upward knowledge distillation, the child node in any pair of parent-child computing nodes acts as the teacher $v^T$, and the parent node acts as the student $v^S$. When performing downward knowledge distillation, the roles are reversed.
%the two nodes perform knowledge distillation as teacher $v^{T}$ and student $v^{S}$ respectively.
Specifically, each model on non-leaf student computing node $Model(v^{S}),\forall v^{S}\notin \mathcal{L}$ distills the knowledge from the model on its teacher $Model(v^{T})$ over the bridge samples that are generated from the embeddings $\varepsilon$ corresponding to $
\bigcup\limits_{u \in Leaf({v^{S}})} {{{\cal D}^u}}  \cap \bigcup\limits_{u \in Leaf({v^{T}})} {{{\cal D}^u}}$.
With generated bridge samples as intermediates, $Model(v^{S})$ is optimized according to the non-leaf distillation loss as follows:
\begin{equation}
	\begin{array}{l}
	\; \; \; \; \mathop {\min }\limits_{{W^{{v^{S}}}}} {L_{non - leaf}}\\ 
	= \mathop {\min }\limits_{{W^{{v^{S}}}}} [{L_{CE}}(\tau (f(dec(\varepsilon );{W^{{v^{S}}}})),{y^\varepsilon }) + \\
	\; \; \; \; \beta  \cdot KL(\tau (f(dec(\varepsilon );{W^{{v^{S}}}}))||\tau (\frac{{{z^\varepsilon }}}{T}))]\\
	 = \mathop {\min }\limits_{{W^{{v^{S}}}}} [{L_{CE}}(\tau (f(dec(\varepsilon );{W^{{v^{S}}}})),{y^\varepsilon }) +
	 \\ \; \; \; \; \beta  \cdot KL(\tau (f(dec(\varepsilon );{W^{{v^{S}}}}))||\tau (\underbrace {\frac{{f(dec(\varepsilon );{W^{{v^{T}}}})}}{T}}_{\frac{{{z^\varepsilon }}}{T}}))],
	\end{array}
	\label{optim-non-leaf}
\end{equation}
where $L_{CE}(\cdot)$ is the cross-entropy loss function, $KL(\cdot)$ is the Kullback-Leibler divergence loss function, and $\beta$ is the distillation weight.
Moreover, $y^\varepsilon$ and $z^\varepsilon$ are the labels and extracted logits of bridge samples corresponding to embeddings $\varepsilon$ uploaded from child nodes, and $\varepsilon$ is ultimately generated at the leaf nodes according to the following equation:
\begin{equation}
    \varepsilon  = enc({X^*}), \forall ({X^*},{y^*}) \in \bigcup\limits_{u \in Leaf({v^S})} {{\mathcal{D}^u}}  \cap \bigcup\limits_{u \in Leaf({v^T})} {{\mathcal{D}^u}}.
    \label{logits-non-leaf}
\end{equation}
Moreover, each model on leaf student computing node $Model(v^{S}),\forall v^{S}\in \mathcal{L}$ is optimized subject to a linear combination of the non-leaf distillation loss over generated bridge samples and the local training losses (controlled by a weight $\gamma$) over private samples $(X^*,y^*)\in \mathcal{D}^{v^{S}}$, that is:
\begin{equation}
		\begin{array}{l}
			 \; \; \; \; \mathop {\min }\limits_{{W^{{v^{S}}}}} {L_{leaf}}\\
			 = \mathop {\min }\limits_{{W^{{v^{S}}}}} {L_{CE}}(f({X^*};{W^{{v^{S}}}}),{y^*}) + \gamma  \cdot {L_{non - leaf}},
		\end{array}
	\label{optim-leaf}
\end{equation}
where
\begin{equation}
	{\quad \varepsilon  = enc({X^*})}, \forall (X^*,y^*)\in \mathcal{D}^{v^{S}}.
    \label{logits-leaf}
\end{equation}
Once the above steps are completed, $v^{S}$ and $v^{T}$ need to swap their roles and optimize in the opposite direction following the above constraints, as shown in Algorithm \ref{alg1}.
Thereout, the models to be trained on every pair of parent-child computing nodes can learn from each other, and the representation learned through knowledge distillation can be propagated to the model on the cloud tier by tier, starting from the leaf nodes.

\subsection{Recursive Agglomeration in End-Edge-Cloud Networks}
%In a two-layer client-server network architecture, the server and clients can directly apply the BSBODP to achieve collaborative training without sharing local data.
Considering FL in an EEC-NET where computing nodes are organized in a tree topology, we propose the FedAgg framework, which applies BSBODP to achieve collaborative training over every pair of parent-child computing nodes, recursively distilling knowledge from bottom to top in an agglomerative manner, as formulated in Algorithm \ref{alg2}. 
Specifically, FedAgg includes two main phases: initialization and recursive training with BSBODP. 
In the initialization phase, leaf nodes generate embeddings using their pre-trained encoders and send them up the tree hierarchy to the root node. 
In the training phase, each node recursively distills the knowledge from the sub-tree that seeks it as the root, and passes the newly extracted knowledge to its parent, applying BSBODP to enable interaction between parent-child pairs.
This process starts from the leaf nodes and ends with the cloud server, where the model on the cloud is updated with the learned knowledge from all tiers. Due to the nature of EEC-NET’s computing resources and the model-agnostic property of BSBODP, the upper-level nodes can deploy larger models and integrate more knowledge from lower-level nodes to achieve superior performance, and the largest model (i.e. the model on the cloud) is exactly the model that we want to gain through collaborative training. 
%It is also worth noting that FedAgg allows parallel training based on BSBODP among multiple parent-child node pairs that are not directly connected via network links, even though they are across different layers.

%which is required by traditional HFL methods. 
%Therefore, FedAgg can reduce synchronization overhead and improve resource utilization across layers.}

%It is worth noting that FedAgg allows parallel training not only between nodes in the same layer, but also across layers, i.e. connected computing nodes close to the cloud and close to data sources can simultaneously communicate and collaboratively train machine learning models according to the BSBODP. In this way, the resource utilization efficiency of the EEC-NET system can be greatly improved compared to previous end-edge-cloud FL methods.
\subsection{Deployment Flexibility Guarantees of Agglomerative Federated Learning}
%\subsubsection{Parallel Collaboration Resilient to Node Failure}
%FedAgg can enhance the resilience to node failure in an EEC-NET, which is a common issue as the number of edge servers increases rapidly. Specifically, FedAgg allows parallel collaboration based on BSBODP among multiple pairs of parent-child nodes that are not directly connected via network links, even if they are across different layers. In this way, FedAgg can avoid the delay caused by waiting for the failed node to recover and immediately start the next communication round without interruption. 
%Moreover, FedAgg can dynamically adjust the aggregation frequency and weight according to the node status and network condition, thus reducing synchronization overhead and improving resource utilization across layers.
%FedAgg can endure node failure to some extent in an EEC-NET, which may occur frequently as the number of edge servers increases rapidly. Specifically, FedAgg enables parallel training based on BSBODP among multiple pairs of parent-child nodes that are not directly connected via network links, even though they are across different layers. In this way, FedAgg can avoid the delay caused by waiting for the failed node to recover and start the next round of aggregation, thus reducing synchronization overhead and improving resource utilization across layers.

%\subsubsection{Dynamic Migration of Computing Nodes}
FedAgg supports the dynamic migration of computing nodes to guarantee deployment flexibility in an EEC-NET, where various factors such as load balancing, connection unreliability, and node failures, may require computing nodes to switch to a different parent at the same level. 
To demonstrate the advantage of FedAgg in handling dynamic migration of computing nodes, we classify existing FL interaction protocols that potentially support hierarchical collaborative model training into two types based on their constraints on model structures of the interacting computing nodes: equivalence interaction protocols and partial order interaction protocols. 
In the following part, we will give formal definitions of these two types of interaction protocols, and prove that equivalence interaction protocols, including our proposed BSBODP adopted by FedAgg, provide better support for dynamic migration of computing nodes.
%, which are equivalence interaction %protocols and partial order interaction protocols.
%In the following part, we will provide formal definitions of the two types of interaction protocols, and prove that equivalent interaction protocols, including BSBODP, can better support the dynamic migration of computing nodes.
%have the benefit of facilitating the dynamic migration of arithmetic nodes.
%In the following part, we will first provide formal definitions,
%of equivalence interaction protocols and partial order protocols, 
%and then prove their support for the dynamic migration of computing nodes.

\noindent
\textbf{Definition 1.} \textit{(Equivalence Interaction Protocol). Consider an EEC-NET with a tree topology formulated in section \ref{eecc}, we define a binary relation $R$ over any pair of parent-child computing nodes.
The sufficient necessary condition for an equivalence interaction protocol is defined as follows:}
\begin{equation}
    < Model({v^i}),Model({v^i}) >  \in R,\forall {v^i} \in \mathcal{V},
\end{equation}
\begin{equation}
    \begin{array}{l}
 < Model({v^i}),Model({v^j}) >  \in R \wedge {v^i} \ne {v^j}\\
 \to  < Model({v^j}),Model({v^i}) >  \in R,\forall {v^i},{v^j} \in \mathcal{V},
\end{array}
\end{equation}
\begin{equation}
\begin{array}{l}
 < Model({v^i}),Model({v^j}) >  \in R\\
 \wedge  < Model({v^j}),Model({v^k}) >  \in R\\
 \to  < Model({v^i}),Model({v^k}) >  \in R,\forall {v^i},{v^j},{v^k} \in \mathcal{V},
\end{array}
\end{equation}
\textit{that is:}
\begin{equation}
   Model(v^i) \sim Model(v^j),\forall <v^i,v^j>\in \mathcal{E}.
\end{equation}
According to our definition, we can easily prove that the followings are equivalent interaction protocols: 1) the same model structure is adopted on the parent and child computing nodes, i.e. $Model(v^i) = Model(v^j),\forall <v^i,v^j>\in \mathcal{E}$, which is represented by FedAvg \cite{mcmahan2017communication}; 
2) model-agnostic interaction protocols that do not impose restrictions on the model structures of parent and child computing nodes, i.e. $Model(v^i) \bot Model(v^j),\forall <v^i,v^j>\in \mathcal{E}$, which is represented by BSBODP.

\noindent
\textbf{Definition 2.} \textit{(Partial Order Interaction Protocol). 
Consider an EEC-NET with a tree topology formulated in section \ref{eecc}, we define a binary relation $R$ over any pair of parent-child computing nodes.
The sufficient necessary condition of a partial order interaction protocol is defined as follows:}
\begin{equation}
    <Model(v^i),Model(v^i)> \in R,\forall{v^i} \in \mathcal{V},
\end{equation}
\begin{equation}
\begin{array}{l}
 < Model({v^i}),Model({v^j}) >  \in R\\
 \wedge  < Model({v^j}),Model({v^i}) >  \in R\\
 \to {v^i} = {v^j},\forall {v^i},{v^j} \in \mathcal{V},
\end{array}
\end{equation}
\begin{equation}
    \begin{array}{l}
 < Model({v^i}),Model({v^j}) >  \in R\\
 \wedge  < Model({v^j}),Model({v^k}) >  \in R\\
 \to  < Model({v^i}),Model({v^k}) >  \in R,\forall {v^i},{v^j},{v^k} \in \mathcal{V},
\end{array}
\end{equation}
\textit{that is:}
\begin{equation}
   Model(v^i) \preceq Model(v^j),\forall <v^i,v^j>\in \mathcal{E}.
\end{equation}
We can also prove that partial training-based interaction protocols \cite{alam2022fedrolex,nguyen2022feddct}, which require the model on the child node to be a sub-model of that on the parent node, i.e. $Model({v^i}) \subseteq Model(Parent({v^i})),\forall  < {v^i},Parent({v^i}) >  \in \mathcal{E}$ are partial order interaction protocols.

\noindent
\textbf{Theorem 1.} \textit{HFL methods based on equivalence interaction protocols allow the parent of any $v^1$ switch to $Parent(v^2)$, where $Parent(v^1)$ and $Parent(v^2)$ are at the same level, i.e. $Model({v^1}) \sim Model(Parent({v^2})),\forall {v^1},{v^2} \in {\mathcal{V}_t},t \in \{ 2,3,...,T\}$}.

\noindent
\textit{Proof.} \\
\noindent
\textbf{Case 1.1.} When ${v^1},{v^2} \in {\mathcal{V}_2}$, we have:
\begin{equation}
    Parent(v^1) = Parent(v^2) = r.
\end{equation}
Hence, $v^1$ and $v^2$ have the same parent node, and there is no dynamic migration of computing nodes in this case.

\noindent
\textbf{Case 1.2.} When ${v^1},{v^2} \in {\mathcal{V}_t},t \ge  3$, we first define 
\begin{equation}
    {Parent^n}(\cdot) = \underbrace {Parent(Parent(\ldots Parent(\cdot)\ldots))}_{{{\rm call} \; Parent( \cdot ) \; {\rm for} \; n \; {\rm times}}},
\end{equation}
then, we have:
\begin{equation}
    \begin{array}{l}
    Model(Parent({v^1})) \sim Model(Paren{t^2}({v^1}))\sim ...\\
    \sim Model(Paren{t^{t - 1}}({v^1})) = r,
    \end{array}
\end{equation}
and 
\begin{equation}
    \begin{array}{l}
Model(Parent({v^2})) \sim Model(Paren{t^2}({v^2}))\sim ...\\
\sim Model(Paren{t^{t - 1}}({v^2})) = r.
\end{array}
\end{equation}
And then,
\begin{equation}
    \begin{array}{l}
    Model({v^1})\sim Model(Parent({v^1}))\\
    \sim r\sim Model(Parent({v^2})),
    \end{array}
\end{equation}
that is:
\begin{equation}
    Model({v^1})\sim Model(Parent({v^2})).
\end{equation}
Therefore, the computing node $v^1$ is allowed to become a child of $Parent(v^2)$, i.e. dynamic migration of computing nodes at the same level is allowed. 
$\hfill \square$

\noindent
\textbf{Theorem 2.} \textit{HFL methods based on partial order interaction protocols do not necessarily allow the parent of any $v^1$ switch to $Parent(v^2)$, where $Parent(v^1)$ and $Parent(v^2)$ are at the same level, i.e. $\neg Model({v^1}) \preceq Model(Parent({v^2})),\exists {v^1},{v^2} \in {\mathcal{V}_t},t \in \{ 2,3,...,T\}$}.

\noindent
\textit{Proof.} \\
\textbf{Case 2.1.} When ${v^1},{v^2} \in {\mathcal{V}_2}$, we have:
\begin{equation}
    Parent(v^1) = Parent(v^2) = r.
\end{equation}
Hence, $v^1$ and $v^2$ have the same parent node $r$, and there is no dynamic migration of computing nodes in this case.

\noindent
\textbf{Case 2.2.} When ${v^1},{v^2} \in {\mathcal{V}_t},t \ge  3$, there are two sub-cases:
\begin{equation}
    Model(Parent(v^1)) \preceq Model(Parent(v^2)),
\end{equation}
and
\begin{equation}
    Model(Parent(v^2)) \preceq Model(Parent(v^1)).
    \label{case2}
\end{equation}
When Eq. (\ref{case2}) is satisfied, there exists a situation where computing node $v^1$ is not allowed to switch its parent to $Parent(v^2)$.
Instantiating $\preceq$ to the partial order relation over integers $\le$, we construct a tree topology $10(9(8,7),5(4,3))$ and set function $Model(\cdot)$ to be a constant function, i.e. $Model(x)=x,\forall x$. 
%In our setting, every two integers satisfy the partial order relation $\le$, 
In our setting, the parent of 7 ($v^1$) and 3 ($v^2$) are at the same level and $Model(Parent(3))=5\le9=Model(Parent(7))$ satisfies Eq. (\ref{case2}). At this point, $\neg Model(7)\le Model(Parent(3))$ is equivalent to $\neg 7\le 5$, which is apparently true. Hence, dynamic migration of computing nodes at the same level is not necessarily allowed.
$\hfill \square$

\section{Experiments}
\subsection{Experiment Setup}
\subsubsection{Experimental Setting} 
Our experiments are conducted based on FedML \cite{he2020fedml}, which is a research library for FL. We use the common CIFAR-10 and CIFAR-100 datasets \cite{cifar10}, which are partitioned into $K$ non-independent and identically distributed parts as private datasets of $K$ different devices. 
The number of clients is set as $K \in \{225, 400\}$ and $K \in \{144,196,225\}$ on CIFAR-10 and CIFAR-100 datasets, respectively.
Following \cite{wu2022exploring}, we adopt a hyper-parameter $\alpha$ pre-set up in FedML to control the degree of data heterogeneity among devices. In our experiments, we set the values of $\alpha \in \{1.0,3.0\}$ corresponding to the high and low degrees of data heterogeneity, respectively.
In terms of network topology, we consider a three-tier tree-structured EEC-NET in our experiments, from bottom to top are end devices, edge nodes, and cloud nodes, where the number of edge servers is set to $\sqrt K$ and only one cloud server is considered. Each end device is either coordinated by an edge server or connected directly to the cloud server.
Regarding model structures, we deploy the same lightweight 6-layer autoencoder pre-trained on ImageNet \cite{deng2009imagenet} on all the end devices and the decoder part of the autoencoder on other computing nodes. In addition, 3-layer-Convolutional Neural Networks (3-layer-CNN), ResNet-10 and ResNet-18 \cite{he2016deep} with small to large model sizes are deployed on the end, edge, and cloud nodes respectively, fitting the capability of computing nodes across different tiers. The configurations of model structures can be found in TABLE \ref{model-setting}.

\subsubsection{Baselines and Criteria}
We compare our proposed FedAgg with four state-of-the-art algorithms, HierFAVG \cite{liu2020client} and DemLearn \cite{nguyen2022self}, which support collaborative model training among multi-tier architectures; FedGKT \cite{he2020group} and FedDKC \cite{wu2022exploring}, which enables training larger models on the server side than that on the client side in a two-tier architecture.
In our experiments, we assume that the servers in FedGKT and FedDKC are deployed on the cloud and directly communicate with end devices. 
In addition, system performance is evaluated by the test accuracy of the cloud-side model after convergence or within 500 communication rounds.

\begin{table}[]
\caption{Model configurations on CIFAR10 dataset. The output dimension of the last fully connected layer needs to be adjusted to 20 on CIFAR100.}
        \setlength{\tabcolsep}{7pt}
	\centering
 \label{model-setting}
 \begin{tabular}{c|c|c|c}
\hline
\textbf{Model Structure}                                            & \textbf{Notation} & \textbf{\begin{tabular}[c]{@{}c@{}}Feature\\ Extractor\end{tabular}} & \textbf{Classifier} \\ \hline
\begin{tabular}[c]{@{}c@{}}3-Layer\\ -CNN\end{tabular}         & $M^{1}$                & \multirow{3}{*}{9.31K}                                               & 10.88K              \\ \cline{1-2} \cline{4-4} 
ResNet-10                                                      & $M^{2}$                &                                                                      & 4.67M               \\ \cline{1-2} \cline{4-4} 
ResNet-18                                                      & $M^{3}$                &                                                                      & 10.65M              \\ \hline
\textbf{Model Structure}                                            & \textbf{Notation} & \textbf{Encoder}                                                     & \textbf{Decoder}    \\ \hline
\begin{tabular}[c]{@{}c@{}}6-Layer\\ -Autoencoder\end{tabular} & -                 & 1.90K                                                                & 2.47K               \\ \hline
\end{tabular}
\end{table}

\begin{table*}[]
\caption{Test accuracy (\%) on CIFAR-10 dataset.}
        \setlength{\tabcolsep}{10pt}
        \centering
\begin{tabular}{l|ccc|cc|cc|c}
\hline
\multicolumn{1}{c|}{\multirow{2}{*}{\textbf{Method}}} & \multicolumn{3}{c|}{\textbf{Model}}                                                                                      & \multicolumn{2}{c|}{\textbf{225 Clients}}                                        & \multicolumn{2}{c|}{\textbf{400 Clients}}                                        & \multirow{2}{*}{\textbf{Average}} \\
\multicolumn{1}{c|}{}                                 & \multicolumn{1}{l}{\textbf{End}}         & \multicolumn{1}{l}{\textbf{Edge}}       & \multicolumn{1}{l|}{\textbf{Cloud}} & \multicolumn{1}{l}{\textbf{$\boldsymbol{\alpha=3.0}$}} & \multicolumn{1}{l|}{\textbf{$\boldsymbol{\alpha=1.0}$}} & \multicolumn{1}{l}{\textbf{$\boldsymbol{\alpha=3.0}$}} & \multicolumn{1}{l|}{\textbf{$\boldsymbol{\alpha=1.0}$}} &                                   \\ \hline
HierFAVG                                             & \multicolumn{1}{c|}{\multirow{5}{*}{$M^1$}} & \multicolumn{2}{c|}{\multirow{2}{*}{$M^1$}}                                      & 10.15                                  & 18.35                                   & 22.25                                  & 18.73                                   & 17.37                             \\ \cline{1-1} \cline{5-9} 
DemLearn                                            & \multicolumn{1}{c|}{}                    & \multicolumn{2}{c|}{}                                                         & 30.55                                  & 32.13                                   & 29.25                                  & 28.06                                   & 30.00                             \\ \cline{1-1} \cline{3-9} 
FedGKT                                               & \multicolumn{1}{c|}{}                    & \multicolumn{1}{c|}{\multirow{2}{*}{-}} & \multirow{3}{*}{$M^3$}                 & 27.39                                  & 26.59                                   & 30.54                                  & 29.22                                   & 28.44                             \\ \cline{1-1} \cline{5-9} 
FedDKC                                            & \multicolumn{1}{c|}{}                    & \multicolumn{1}{c|}{}                   &                                     & 32.43                                  & 30.09                                   & 30.79                                  & 28.78                                   & 30.52                             \\ \cline{1-1} \cline{3-3} \cline{5-9} 
\textbf{FedAgg}                                       & \multicolumn{1}{c|}{}                    & \multicolumn{1}{c|}{$M^2$}                 &                                     & \textbf{36.76}                         & \textbf{33.87}                          & \textbf{35.83}                         & \textbf{33.74}                          & \textbf{35.05}                    \\ \hline
\end{tabular}
\label{result-cifar10}
%\end{table*}

\vspace{13pt}

%\begin{table*}[]
\caption{Test accuracy (\%) on CIFAR-100 dataset, taking $\alpha=3.0$.}
                \setlength{\tabcolsep}{9pt}
        \centering
\begin{tabular}{l|ccc|c|c|c|c}
\hline
\multicolumn{1}{c|}{\multirow{2}{*}{\textbf{Method}}} & \multicolumn{3}{c|}{\textbf{Model}}                                                           & \multirow{2}{*}{\textbf{144 Clients}} & \multirow{2}{*}{\textbf{196 Clients}} & \multirow{2}{*}{\textbf{225 Clients}} & \multirow{2}{*}{\textbf{Average}} \\
\multicolumn{1}{c|}{}                                 & \multicolumn{1}{c}{\textbf{End}}         & \multicolumn{1}{c}{\textbf{Edge}} & \textbf{Cloud} &                                       &                                       &                                       &                                   \\ \hline
HierFAVG                                              & \multicolumn{1}{c|}{\multirow{6}{*}{$M^1$}} & \multicolumn{2}{c|}{\multirow{5}{*}{$M^1$}}           & 9.11                                  & 9.42                                  & 5.03                                  & 7.85                              \\ \cline{1-1} \cline{5-8} 
DemLearn ($lv$=3)                                         & \multicolumn{1}{c|}{}                    & \multicolumn{2}{c|}{}                              & 15.75                                 & 14.94                                 & 14.81                                 & 15.17                             \\ \cline{1-1} \cline{5-8} 
DemLearn ($lv$=5)                                         & \multicolumn{1}{c|}{}                    & \multicolumn{2}{c|}{}                              & 14.99                                 & 14.59                                 & 14.55                                 & 14.71                             \\ \cline{1-1} \cline{5-8} 
DemLearn ($lv$=7)                                         & \multicolumn{1}{c|}{}                    & \multicolumn{2}{c|}{}                              & 14.52                                 & 13.97                                 & 13.89                                 & 14.13                             \\ \cline{1-1} \cline{5-8} 
DemLearn ($lv$=10)                                        & \multicolumn{1}{c|}{}                    & \multicolumn{2}{c|}{}                              & 12.56                                 & 13.25                                 & 12.52                                 & 12.78                             \\ \cline{1-1} \cline{3-8} 
\textbf{FedAgg}                                       & \multicolumn{1}{c|}{}                    & \multicolumn{1}{c|}{$M^2$}           & $M^3$             & \textbf{17.07}                        & \textbf{17.33}                        & \textbf{16.50}                        & \textbf{16.97}                    \\ \hline
\end{tabular}
\label{result-cifar100}
\end{table*}

\begin{table}[]
\caption{Communication rounds when reaching a given test accuracy on CIFAR-10 dataset. Communication rounds are counted from 0, the same as below.}
                \setlength{\tabcolsep}{2.5pt}
        \centering
\begin{tabular}{l|cccc|cccc}
\hline
\multicolumn{1}{c|}{\multirow{3}{*}{\textbf{Method}}} & \multicolumn{4}{c|}{\textbf{225 Clients}}                                & \multicolumn{4}{c}{\textbf{400 Clients}}                                \\
\multicolumn{1}{c|}{}                                 & \multicolumn{2}{c}{\textbf{$\boldsymbol{\alpha=3.0}$}} & \multicolumn{2}{c|}{\textbf{$\boldsymbol{\alpha=1.0}$}} & \multicolumn{2}{c}{\textbf{$\boldsymbol{\alpha=3.0}$}} & \multicolumn{2}{c}{\textbf{$\boldsymbol{\alpha=1.0}$}} \\
\multicolumn{1}{c|}{}                                 & \textbf{25\%}    & \textbf{30\%}   & \textbf{25\%}    & \textbf{30\%}    & \textbf{25\%}    & \textbf{30\%}   & \textbf{25\%}    & \textbf{30\%}   \\ \hline
HierFAVG                                              & -                & -               & -                & -                & -                & -               & -                & -               \\ \hline
DemLearn                                              & 138              & 458             & 113              & 333              & 216              & -               & 243              & -               \\ \hline
FedGKT                                                & 12               & -               & 23               & -                & 6                & 18              & 11               & -               \\ \hline
FedDKC                                                & 8                & 14              & 12               & 23               & 7                & 24              & 12               & -               \\ \hline
\textbf{FedAgg}                                       & \textbf{0}       & \textbf{2}      & \textbf{1}       & \textbf{7}       & \textbf{1}       & \textbf{4}      & \textbf{5}       & \textbf{10}     \\ \hline
\end{tabular}
\label{conv-cifar-10}
\end{table}

\subsubsection{Hyperparameters} 
We conduct the learning rate of 0.001 with batch size 8 to all methods. Besides, personalized hyper-parameters are set as follows:
\begin{itemize}
    \item 
    HierFAVG adopts $\kappa_1=1$ and $\kappa_2=1$.
    \item 
    DemLearn adopts the default hyper-parameters setting in \cite{demlearn-code}.
    \item 
    FedGKT adopts $\beta=1.5$.
    \item 
    FedDKC adopts KKR as the knowledge refinement strategy with $\beta=1.5$ and $T=0.12$.
    %\item 
    %FedGEMS adopt ???, following XXX
    \item 
    FedAgg adopt $\gamma=1$, $T=3$ and $\beta=10$.
\end{itemize}

\begin{figure}[!t]
	\centering
	\includegraphics[width=0.46\textwidth]{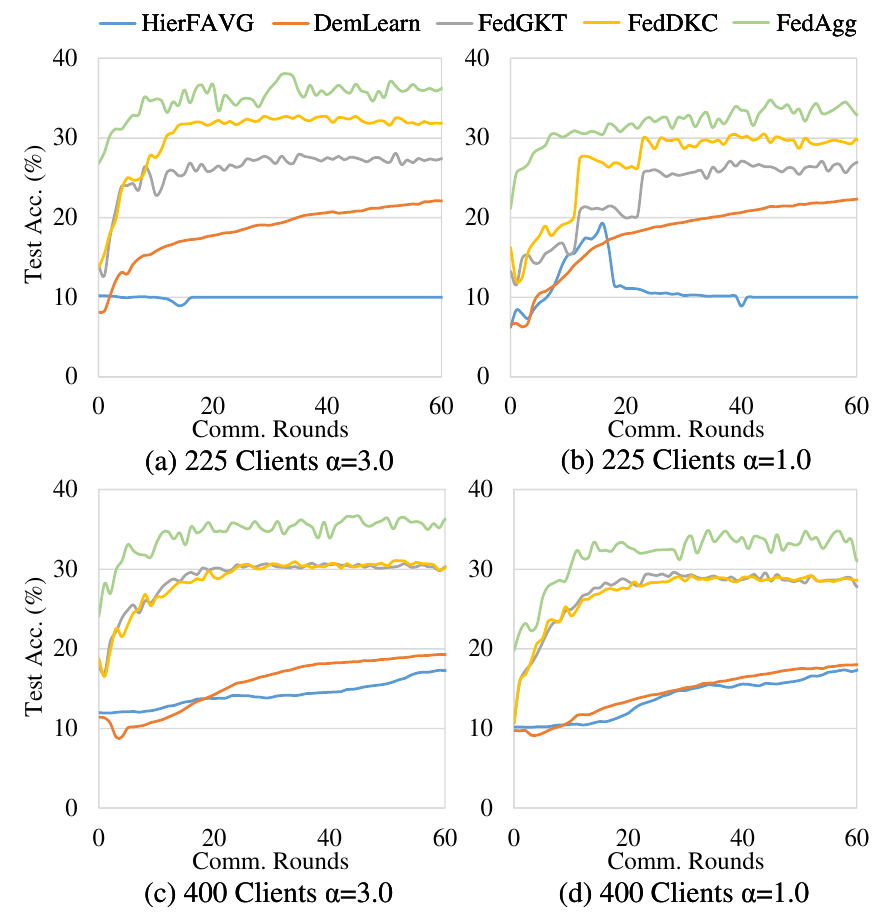}
	\caption{Learning curves of the cloud-side model with varying numbers of devices and degrees of data heterogeneity. Results are obtained on CIFAR-10 dataset.}
	\label{curve1}
%\end{figure}

\vspace{10pt}

%\begin{figure}[t]
	\centering
	\includegraphics[width=0.46\textwidth]{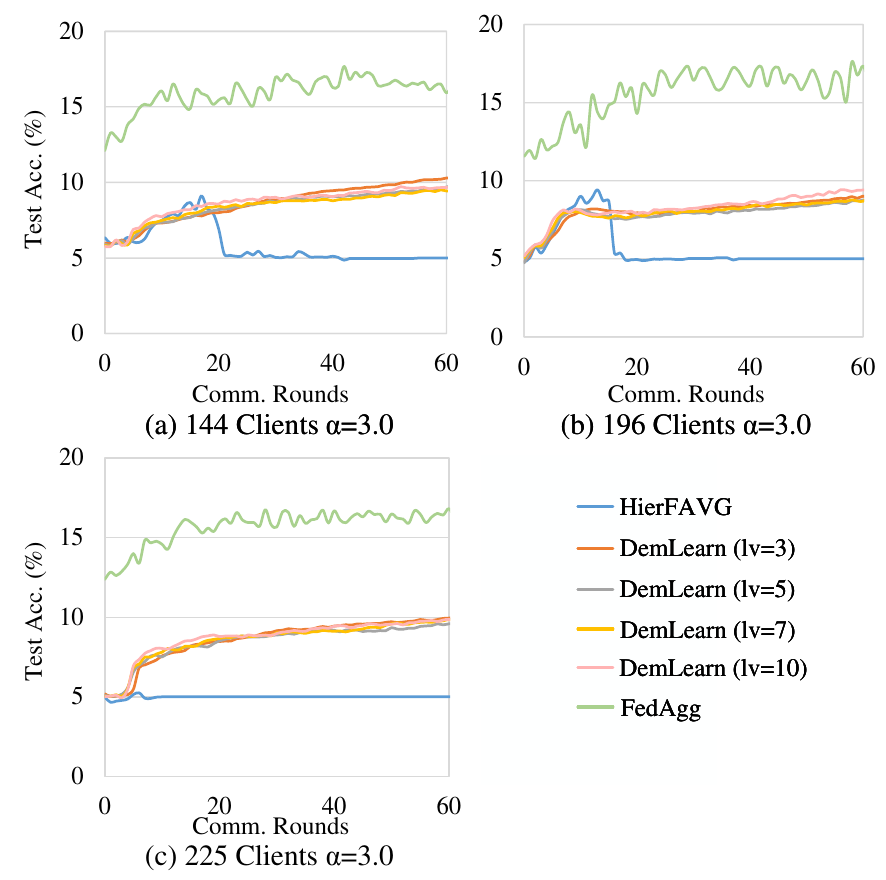}
	\caption{Learning curves of the cloud-side model with varying numbers of devices. Results are obtained on CIFAR-100 dataset.}
	\label{curve2}
\end{figure}

\subsection{Performance Evaluation}
\subsubsection{Comparison of Test Accuracy}
We compare FedAgg with HierFAVG, DemLearn, FedGKT, and FedDKC under different numbers of clients and data heterogeneity settings on CIFAR-10, and the results are shown in TABLE \ref{result-cifar10}. We can see that FedAgg achieves the highest accuracy in all settings, which outperforms the highest accuracy among HierFAVG, DemLearn, FedGKT, and FedDKC by 4.33\%, 1.74\%, 5.04\% and 4.52\% in experimental settings from left to right.
These results not only show the better performance of FedAgg, but also demonstrate its ability to handle different data distributions and the varying number of clients.
To further confirm the superiority of FedAgg under different organizing structures of clients, we compare the performance of FedAgg with HierFAVG and DemLearn on CIFAR-100 under varying numbers of hierarchical structure levels ($lv \in \{3,5,7,10\}$) adopted in DemLearn. As shown in TABLE \ref{result-cifar100}, FedAgg outperforms the baseline algorithms in all experimental settings over CIFAR-100, which outperforms the highest accuracy among HierFAVG and DemLearn with varying hierarchical structures by 1.32\%, 2.39\%, 1.69\% respectively in terms of 144, 196 and 225 clients.
This demonstrates that FedAgg can benefit from exploiting the potential of deploying larger models on edge and cloud nodes in an EEC-NET, and is suitable for scenarios with different numbers of devices.
%This demonstrates that FedAgg can fully exploit deploying large model potential in an EEC-NET to achieve superior performance, and can also adapt to diverse numbers of devices.

\subsubsection{Convergence Rate}
We evaluate the convergence rate of FedAgg against baseline algorithms on both CIFAR-10 and CIFAR-100 datasets from two perspectives: the required communication rounds when reaching a given test accuracy, and the learning curves. 
TABLE \ref{conv-cifar-10} and \ref{conv-cifar-100} show the communication rounds needed to reach a given test accuracy on CIFAR-10 and CIFAR-100 datasets, respectively. As shown in the two tables, FedAgg requires much fewer communication rounds to converge to a given test accuracy than all compared algorithms, thus demonstrating remarkable improvements in convergence rate. 
Besides, Fig. \ref{curve1} shows the learning curves under different degrees of data heterogeneity and varying numbers of clients on CIFAR-10, while Fig. \ref{curve2} shows the learning curves under different organizing structures on CIFAR-100. 
From Fig. \ref{curve1} and Fig. \ref{curve2}, we can conclude that FedAgg consistently outperforms the baseline algorithms in terms of test accuracy with the same number of communication rounds, and also guarantees faster convergence under various experimental settings.

\begin{table}[]
\caption{Communication rounds when reaching a given test accuracy on CIFAR-100 dataset, taking $\alpha=3.0$.}
                \setlength{\tabcolsep}{3pt}
        \centering
\begin{tabular}{l|cc|cc|cc}
\hline
\multicolumn{1}{c|}{\multirow{2}{*}{\textbf{Method}}} & \multicolumn{2}{c|}{\textbf{144 Clients}} & \multicolumn{2}{c|}{\textbf{196 Clients}} & \multicolumn{2}{c}{\textbf{225 Clients}} \\
\multicolumn{1}{c|}{}                                 & \textbf{9\%}        & \textbf{14\%}       & \textbf{9\%}        & \textbf{14\%}       & \textbf{9\%}       & \textbf{14\%}       \\ \hline
HierFAVG                                              & 17                  & -                   & 13                  & -                   & -                  & -                   \\ \hline
DemLearn ($lv$=3)                                       & 33                  & 266                 & 64                  & 365                 & 28                 & 350                 \\ \hline
DemLearn ($lv$=5)                                       & 36                  & 347                 & 69                  & 386                 & 34                 & 436                 \\ \hline
DemLearn ($lv$=7)                                       & 46                  & 425                 & 64                  & -                   & 30                 & -                   \\ \hline
DemLearn ($lv$=10)                                      & 33                  & -                   & 50                  & -                   & 31                 & -                   \\ \hline
\textbf{FedAgg}                                       & \textbf{0}          & \textbf{5}          & \textbf{0}          & \textbf{12}         & \textbf{0}         & \textbf{7}          \\ \hline
\end{tabular}
\label{conv-cifar-100}
\end{table}

\section{Ablation Study}
\subsection{Effectiveness of Online Distillation}
TABLE \ref{ablation-beta} shows the impact of the distillation weight $\beta$ on the performance of the cloud-side model. 
We can observe that when $\beta$ is set to a relatively small value ($\beta=0.1$), the performance of the cloud-side model is poor, which indicates the effectiveness of transferring knowledge from lower-level computing nodes through online distillation.
However, when $\beta$ is set to a very large value ($\beta=50$), the performance of the cloud-side model drops significantly due to the excessive weakening of cross-entropy-based optimization.

\begin{table}[]
\caption{FedAgg with different $\beta$. Experiments are conducted on CIFAR-10 dataset with 225 clients, taking $\alpha=1.0$.}
        \setlength{\tabcolsep}{3.6pt}
	\centering
\begin{tabular}{l|ccccc}
\hline
\multicolumn{1}{c|}{\textbf{Method}} & \textbf{$\boldsymbol{\beta=0.1}$} & \textbf{$\boldsymbol{\beta=1}$} & \textbf{$\boldsymbol{\beta=3}$} & \textbf{$\boldsymbol{\beta=10}$} & \textbf{$\boldsymbol{\beta=50}$} \\ \hline
\textbf{FedAgg}                      & 33.14          & \textbf{35.33}        & 34.66        & 33.87         & 27.21         \\ \hline
\end{tabular}
\label{ablation-beta}
%\end{table}

\vspace{10pt}

%\begin{table}[]
\caption{FedAgg with different model heterogeneity setup on devices. Experiments are conducted on CIFAR-10 dataset with 225 clients, taking $\alpha=1.0$.}
        \setlength{\tabcolsep}{9pt}
	\centering
 \label{ablation-device}
\begin{tabular}{l|cc}
\hline
\multicolumn{1}{c|}{\textbf{Method}} & \textbf{Model Homo.} & \textbf{Model Hetero.} \\ \hline
\textbf{FedAgg}                      & \textbf{33.87}                 & 32.83                   \\ \hline
\end{tabular}
%\end{table}

\vspace{10pt}

%\begin{table}[]
\caption{FedAgg with different model structures on the cloud. Experiments are conducted on CIFAR-10 dataset with 225 clients, taking $\alpha=1.0$.}
        \setlength{\tabcolsep}{7pt}
	\centering
 \label{ablation-model}
\begin{tabular}{l|ccc}
\hline
\multicolumn{1}{c|}{\textbf{Method}} & \textbf{3-Layer-CNN} & \textbf{ResNet-10} & \textbf{ResNet-18} \\ \hline
\textbf{FedAgg}                      & 23.84                & 31.11              & \textbf{33.87}              \\ \hline
\end{tabular}
\end{table}

\subsection{Tolerance to Device Heterogeneity}
%To analyze model heterogeneity on devices, we change the model structure of 20\% clients from $M^1$ to $M^2$, and compare their experimental results with those of all clients with the same model structure $M^1$, as shown in TABLE \ref{ablation-device}. 
%The results show that the change of the model structures on clients has little effect on the performance of the cloud-side model, indicating that FedAgg is robust to heterogeneous device-side models and is suitable to be applied in the EECC scenario with capability-differentiated computing nodes.
To examine the impact of heterogeneous on-device models on the system performance, we modify the model structure of 20\% clients from $M^1$ to $M^2$, and compare their experimental results with those of all clients using the same model structure $M^1$, as shown in TABLE \ref{ablation-device}.
The results demonstrate that the variation of the model structures on clients has a negligible effect on the performance of the cloud-side model. 
This implies that FedAgg can tolerate the heterogeneity of device-side models and is suitable for the EECC scenario where devices have different computing capabilities.

\subsection{Smaller Cloud-side Models}
We compare the performance of FedAgg using smaller cloud-side models $M^1$ and $M^2$ with that obtained by using a larger cloud model $M^3$, as shown in TABLE \ref{ablation-model}. 
We observe that the performance of FedAgg drops significantly as the model size on the cloud decreases.
This further confirms that our proposed FedAgg can achieve remarkable performance gains by enabling the deployment of larger models on the cloud than on end devices.

\section{Conclusion}
To overcome the limitation of model scale on powerful edge and cloud computing nodes constrained by the weakest end devices in EECC-empowered FL, we propose a novel federated learning framework called Agglomerative Federated Learning (FedAgg), which allows the trained models from end devices, bridge edge servers, to cloud servers to grow larger in size and stronger in generalization ability.
To be specific, FedAgg recursively distills knowledge from bottom to top in an agglomeration manner via our customized Bridge Sample Based Online Distillation Protocol (BSBODP), which can achieve model-agnostic interaction across every pair of parent-child computing nodes through online distillation over generated bridge samples. To our best knowledge, FedAgg is the first framework empowered by EECC that enables training larger models with ever-increasing capability tier by tier up to the cloud, and can achieve superior performance than related state-of-the-art methods in terms of test accuracy and convergence rate.

\section*{Acknowledgment}
This work was supported by the National Key Research and Development Program of China (No. 2021YFB2900102), the National Natural Science Foundation of China (No. 62072436), the Innovation Capability Support Program of Shaanxi (No. 2023-CX-TD-08), Shaanxi Qinchuangyuan "scientists+engineers" team (No. 2023KXJ-040), the Innovation Funding of ICT, CAS (No. E261080). We thank Zeju Li from Beijing University of Posts and Telecommunications, Sijie Cheng from Tsinghua University, Tian Wen, Wen Wang, and Yufeng Chen from Institute of Computing Technology, Chinese Academy of Sciences, and Jinda Lu from the University of Science and Technology of China for inspiring suggestions.

\bibliographystyle{IEEEtran}

% Generated by IEEEtran.bst, version: 1.14 (2015/08/26)
\begin{thebibliography}{10}
\providecommand{\url}[1]{#1}
\csname url@samestyle\endcsname
\providecommand{\newblock}{\relax}
\providecommand{\bibinfo}[2]{#2}
\providecommand{\BIBentrySTDinterwordspacing}{\spaceskip=0pt\relax}
\providecommand{\BIBentryALTinterwordstretchfactor}{4}
\providecommand{\BIBentryALTinterwordspacing}{\spaceskip=\fontdimen2\font plus
\BIBentryALTinterwordstretchfactor\fontdimen3\font minus
  \fontdimen4\font\relax}
\providecommand{\BIBforeignlanguage}[2]{{%
\expandafter\ifx\csname l@#1\endcsname\relax
\typeout{** WARNING: IEEEtran.bst: No hyphenation pattern has been}%
\typeout{** loaded for the language `#1'. Using the pattern for}%
\typeout{** the default language instead.}%
\else
\language=\csname l@#1\endcsname
\fi
#2}}
\providecommand{\BIBdecl}{\relax}
\BIBdecl

\bibitem{li-rtgwg-cfn-framework-00}
\BIBentryALTinterwordspacing
Y.~Li, J.~HE, L.~Geng, P.~Liu, and Y.~Cui, ``{Framework of Compute First
  Networking (CFN)},'' Internet Engineering Task Force, Internet-Draft
  draft-li-rtgwg-cfn-framework-00, Nov. 2019, work in Progress. [Online].
  Available:
  \url{https://datatracker.ietf.org/doc/draft-li-rtgwg-cfn-framework/00/}
\BIBentrySTDinterwordspacing

\bibitem{krol2019compute}
M.~Kr{\'o}l, S.~Mastorakis, D.~Oran, and D.~Kutscher, ``Compute first
  networking: Distributed computing meets icn,'' in \emph{Proceedings of the
  6th ACM Conference on Information-Centric Networking}, 2019, pp. 67--77.

\bibitem{tian2021overview}
L.~Tian, M.~Yang, and S.~Wang, ``An overview of compute first networking,''
  \emph{International Journal of Web and Grid Services}, vol.~17, no.~2, pp.
  81--97, 2021.

\bibitem{yang2019federated}
Q.~Yang, Y.~Liu, T.~Chen, and Y.~Tong, ``Federated machine learning: Concept
  and applications,'' \emph{ACM Transactions on Intelligent Systems and
  Technology (TIST)}, vol.~10, no.~2, pp. 1--19, 2019.

\bibitem{lim2020federated}
W.~Y.~B. Lim, N.~C. Luong, D.~T. Hoang, Y.~Jiao, Y.-C. Liang, Q.~Yang,
  D.~Niyato, and C.~Miao, ``Federated learning in mobile edge networks: A
  comprehensive survey,'' \emph{IEEE Communications Surveys \& Tutorials},
  vol.~22, no.~3, pp. 2031--2063, 2020.

\bibitem{kairouz2021advances}
P.~Kairouz, H.~B. McMahan, B.~Avent, A.~Bellet, M.~Bennis, A.~N. Bhagoji,
  K.~Bonawitz, Z.~Charles, G.~Cormode, R.~Cummings \emph{et~al.}, ``Advances
  and open problems in federated learning,'' \emph{Foundations and
  Trends{\textregistered} in Machine Learning}, vol.~14, no. 1--2, pp. 1--210,
  2021.

\bibitem{mcmahan2017communication}
B.~McMahan, E.~Moore, D.~Ramage, S.~Hampson, and B.~A. y~Arcas,
  ``Communication-efficient learning of deep networks from decentralized
  data,'' in \emph{Artificial intelligence and statistics}.\hskip 1em plus
  0.5em minus 0.4em\relax PMLR, 2017, pp. 1273--1282.

\bibitem{li2020federated}
T.~Li, A.~K. Sahu, M.~Zaheer, M.~Sanjabi, A.~Talwalkar, and V.~Smith,
  ``Federated optimization in heterogeneous networks,'' \emph{Proceedings of
  Machine Learning and Systems}, vol.~2, pp. 429--450, 2020.

\bibitem{karimireddy2020scaffold}
S.~P. Karimireddy, S.~Kale, M.~Mohri, S.~Reddi, S.~Stich, and A.~T. Suresh,
  ``Scaffold: Stochastic controlled averaging for federated learning,'' in
  \emph{International Conference on Machine Learning}.\hskip 1em plus 0.5em
  minus 0.4em\relax PMLR, 2020, pp. 5132--5143.

\bibitem{liu2020client}
L.~Liu, J.~Zhang, S.~Song, and K.~B. Letaief, ``Client-edge-cloud hierarchical
  federated learning,'' in \emph{ICC 2020-2020 IEEE International Conference on
  Communications (ICC)}.\hskip 1em plus 0.5em minus 0.4em\relax IEEE, 2020, pp.
  1--6.

\bibitem{wang2022accelerating}
Z.~Wang, H.~Xu, J.~Liu, Y.~Xu, H.~Huang, and Y.~Zhao, ``Accelerating federated
  learning with cluster construction and hierarchical aggregation,'' \emph{IEEE
  Transactions on Mobile Computing}, 2022.

\bibitem{wang2021resource}
Z.~Wang, H.~Xu, J.~Liu, H.~Huang, C.~Qiao, and Y.~Zhao, ``Resource-efficient
  federated learning with hierarchical aggregation in edge computing,'' in
  \emph{IEEE INFOCOM 2021-IEEE Conference on Computer Communications}.\hskip
  1em plus 0.5em minus 0.4em\relax IEEE, 2021, pp. 1--10.

\bibitem{feng2022mobility}
C.~Feng, H.~H. Yang, D.~Hu, Z.~Zhao, T.~Q. Quek, and G.~Min, ``Mobility-aware
  cluster federated learning in hierarchical wireless networks,'' \emph{IEEE
  Transactions on Wireless Communications}, vol.~21, no.~10, pp. 8441--8458,
  2022.

\bibitem{nguyen2022self}
M.~N. Nguyen, S.~R. Pandey, T.~N. Dang, E.-N. Huh, N.~H. Tran, W.~Saad, and
  C.~S. Hong, ``Self-organizing democratized learning: Toward large-scale
  distributed learning systems,'' \emph{IEEE Transactions on Neural Networks
  and Learning Systems}, 2022.

\bibitem{nguyen2021distributed}
M.~N. Nguyen, S.~R. Pandey, K.~Thar, N.~H. Tran, M.~Chen, W.~S. Bradley, and
  C.~S. Hong, ``Distributed and democratized learning: Philosophy and research
  challenges,'' \emph{IEEE Computational Intelligence Magazine}, vol.~16,
  no.~1, pp. 49--62, 2021.

\bibitem{hu2021model}
X.~Hu, L.~Chu, J.~Pei, W.~Liu, and J.~Bian, ``Model complexity of deep
  learning: A survey,'' \emph{Knowledge and Information Systems}, vol.~63, pp.
  2585--2619, 2021.

\bibitem{alam2022fedrolex}
S.~Alam, L.~Liu, M.~Yan, and M.~Zhang, ``Fedrolex: Model-heterogeneous
  federated learning with rolling sub-model extraction,'' in \emph{Conference
  on Neural Information Processing Systems (NeurIPS)}, 2022.

\bibitem{nguyen2022feddct}
Q.~Nguyen, H.~H. Pham, K.-S. Wong, P.~L. Nguyen, T.~T. Nguyen, and M.~N. Do,
  ``Feddct: Federated learning of large convolutional neural networks on
  resource constrained devices using divide and co-training,'' \emph{arXiv
  preprint arXiv:2211.10948}, 2022.

\bibitem{cho2022heterogeneous}
\BIBentryALTinterwordspacing
Y.~J. Cho, A.~Manoel, G.~Joshi, R.~Sim, and D.~Dimitriadis, ``Heterogeneous
  ensemble knowledge transfer for training large models in federated
  learning,'' in \emph{Proceedings of the Thirty-First International Joint
  Conference on Artificial Intelligence, {IJCAI-22}}, L.~D. Raedt, Ed.\hskip
  1em plus 0.5em minus 0.4em\relax International Joint Conferences on
  Artificial Intelligence Organization, 7 2022, pp. 2881--2887, main Track.
  [Online]. Available: \url{https://doi.org/10.24963/ijcai.2022/399}
\BIBentrySTDinterwordspacing

\bibitem{cheng2021fedgems}
S.~Cheng, J.~Wu, Y.~Xiao, and Y.~Liu, ``Fedgems: Federated learning of larger
  server models via selective knowledge fusion,'' \emph{arXiv preprint
  arXiv:2110.11027}, 2021.

\bibitem{he2020group}
C.~He, M.~Annavaram, and S.~Avestimehr, ``Group knowledge transfer: Federated
  learning of large cnns at the edge,'' \emph{Advances in Neural Information
  Processing Systems}, vol.~33, pp. 14\,068--14\,080, 2020.

\bibitem{tak2020federated}
A.~Tak and S.~Cherkaoui, ``Federated edge learning: Design issues and
  challenges,'' \emph{IEEE Network}, vol.~35, no.~2, pp. 252--258, 2020.

\bibitem{anil2018large}
R.~Anil, G.~Pereyra, A.~Passos, R.~Ormandi, G.~E. Dahl, and G.~E. Hinton,
  ``Large scale distributed neural network training through online
  distillation,'' \emph{arXiv preprint arXiv:1804.03235}, 2018.

\end{thebibliography}


\begin{thebibliography}{10}
\providecommand{\url}[1]{#1}
\csname url@samestyle\endcsname
\providecommand{\newblock}{\relax}
\providecommand{\bibinfo}[2]{#2}
\providecommand{\BIBentrySTDinterwordspacing}{\spaceskip=0pt\relax}
\providecommand{\BIBentryALTinterwordstretchfactor}{4}
\providecommand{\BIBentryALTinterwordspacing}{\spaceskip=\fontdimen2\font plus
\BIBentryALTinterwordstretchfactor\fontdimen3\font minus
  \fontdimen4\font\relax}
\providecommand{\BIBforeignlanguage}[2]{{%
\expandafter\ifx\csname l@#1\endcsname\relax
\typeout{** WARNING: IEEEtran.bst: No hyphenation pattern has been}%
\typeout{** loaded for the language `#1'. Using the pattern for}%
\typeout{** the default language instead.}%
\else
\language=\csname l@#1\endcsname
\fi
#2}}
\providecommand{\BIBdecl}{\relax}
\BIBdecl

\bibitem{kairouz2021advances}
P.~Kairouz, H.~B. McMahan, B.~Avent, A.~Bellet, M.~Bennis, A.~N. Bhagoji,
  K.~Bonawitz, Z.~Charles, G.~Cormode, R.~Cummings \emph{et~al.}, ``Advances
  and open problems in federated learning,'' \emph{Foundations and
  Trends{\textregistered} in Machine Learning}, vol.~14, no. 1--2, pp. 1--210,
  2021.

\bibitem{yang2019federated}
Q.~Yang, Y.~Liu, T.~Chen, and Y.~Tong, ``Federated machine learning: Concept
  and applications,'' \emph{ACM Transactions on Intelligent Systems and
  Technology (TIST)}, vol.~10, no.~2, pp. 1--19, 2019.

\bibitem{nguyen2021federated}
D.~C. Nguyen, M.~Ding, P.~N. Pathirana, A.~Seneviratne, J.~Li, and H.~V. Poor,
  ``Federated learning for internet of things: A comprehensive survey,''
  \emph{IEEE Communications Surveys \& Tutorials}, vol.~23, no.~3, pp.
  1622--1658, 2021.

\bibitem{fliotsurvey2}
T.~Zhang, L.~Gao, C.~He, M.~Zhang, B.~Krishnamachari, and A.~S. Avestimehr,
  ``Federated learning for the internet of things: Applications, challenges,
  and opportunities,'' \emph{IEEE Internet of Things Magazine}, vol.~5, no.~1,
  pp. 24--29, 2022.

\bibitem{duan2022distributed}
S.~Duan, D.~Wang, J.~Ren, F.~Lyu, Y.~Zhang, H.~Wu, and X.~Shen, ``Distributed
  artificial intelligence empowered by end-edge-cloud computing: A survey,''
  \emph{IEEE Communications Surveys \& Tutorials}, 2022.

\bibitem{bao2022federated}
G.~Bao and P.~Guo, ``Federated learning in cloud-edge collaborative
  architecture: key technologies, applications and challenges,'' \emph{Journal
  of Cloud Computing}, vol.~11, no.~1, p.~94, 2022.

\bibitem{mcmahan2017communication}
B.~McMahan, E.~Moore, D.~Ramage, S.~Hampson, and B.~A. y~Arcas,
  ``Communication-efficient learning of deep networks from decentralized
  data,'' in \emph{Artificial intelligence and statistics}.\hskip 1em plus
  0.5em minus 0.4em\relax PMLR, 2017, pp. 1273--1282.

\bibitem{li2020federated}
T.~Li, A.~K. Sahu, M.~Zaheer, M.~Sanjabi, A.~Talwalkar, and V.~Smith,
  ``Federated optimization in heterogeneous networks,'' \emph{Proceedings of
  Machine Learning and Systems}, vol.~2, pp. 429--450, 2020.

\bibitem{karimireddy2020scaffold}
S.~P. Karimireddy, S.~Kale, M.~Mohri, S.~Reddi, S.~Stich, and A.~T. Suresh,
  ``Scaffold: Stochastic controlled averaging for federated learning,'' in
  \emph{International Conference on Machine Learning}.\hskip 1em plus 0.5em
  minus 0.4em\relax PMLR, 2020, pp. 5132--5143.

\bibitem{liu2020client}
L.~Liu, J.~Zhang, S.~Song, and K.~B. Letaief, ``Client-edge-cloud hierarchical
  federated learning,'' in \emph{ICC 2020-2020 IEEE International Conference on
  Communications (ICC)}.\hskip 1em plus 0.5em minus 0.4em\relax IEEE, 2020, pp.
  1--6.

\bibitem{tak2020federated}
A.~Tak and S.~Cherkaoui, ``Federated edge learning: Design issues and
  challenges,'' \emph{IEEE Network}, vol.~35, no.~2, pp. 252--258, 2020.

\bibitem{wang2022accelerating}
Z.~Wang, H.~Xu, J.~Liu, Y.~Xu, H.~Huang, and Y.~Zhao, ``Accelerating federated
  learning with cluster construction and hierarchical aggregation,'' \emph{IEEE
  Transactions on Mobile Computing}, 2022.

\bibitem{wang2021resource}
Z.~Wang, H.~Xu, J.~Liu, H.~Huang, C.~Qiao, and Y.~Zhao, ``Resource-efficient
  federated learning with hierarchical aggregation in edge computing,'' in
  \emph{IEEE INFOCOM 2021-IEEE Conference on Computer Communications}.\hskip
  1em plus 0.5em minus 0.4em\relax IEEE, 2021, pp. 1--10.

\bibitem{feng2022mobility}
C.~Feng, H.~H. Yang, D.~Hu, Z.~Zhao, T.~Q. Quek, and G.~Min, ``Mobility-aware
  cluster federated learning in hierarchical wireless networks,'' \emph{IEEE
  Transactions on Wireless Communications}, vol.~21, no.~10, pp. 8441--8458,
  2022.

\bibitem{makd}
\BIBentryALTinterwordspacing
A.~Afonin and S.~P. Karimireddy, ``Towards model agnostic federated learning
  using knowledge distillation,'' \emph{CoRR}, vol. abs/2110.15210, 2021.
  [Online]. Available: \url{https://arxiv.org/abs/2110.15210}
\BIBentrySTDinterwordspacing

\bibitem{nguyen2022self}
M.~N. Nguyen, S.~R. Pandey, T.~N. Dang, E.-N. Huh, N.~H. Tran, W.~Saad, and
  C.~S. Hong, ``Self-organizing democratized learning: Toward large-scale
  distributed learning systems,'' \emph{IEEE Transactions on Neural Networks
  and Learning Systems}, 2022.

\bibitem{nguyen2021distributed}
M.~N. Nguyen, S.~R. Pandey, K.~Thar, N.~H. Tran, M.~Chen, W.~S. Bradley, and
  C.~S. Hong, ``Distributed and democratized learning: Philosophy and research
  challenges,'' \emph{IEEE Computational Intelligence Magazine}, vol.~16,
  no.~1, pp. 49--62, 2021.

\bibitem{hu2021model}
X.~Hu, L.~Chu, J.~Pei, W.~Liu, and J.~Bian, ``Model complexity of deep
  learning: A survey,'' \emph{Knowledge and Information Systems}, vol.~63, pp.
  2585--2619, 2021.

\bibitem{alam2022fedrolex}
S.~Alam, L.~Liu, M.~Yan, and M.~Zhang, ``Fedrolex: Model-heterogeneous
  federated learning with rolling sub-model extraction,'' in \emph{Conference
  on Neural Information Processing Systems (NeurIPS)}, 2022.

\bibitem{nguyen2022feddct}
Q.~Nguyen, H.~H. Pham, K.-S. Wong, P.~Le~Nguyen, T.~T. Nguyen, and M.~N. Do,
  ``Feddct: Federated learning of large convolutional neural networks on
  resource constrained devices using divide and collaborative training,''
  \emph{IEEE Transactions on Network and Service Management}, 2023.

\bibitem{cho2022heterogeneous}
\BIBentryALTinterwordspacing
Y.~J. Cho, A.~Manoel, G.~Joshi, R.~Sim, and D.~Dimitriadis, ``Heterogeneous
  ensemble knowledge transfer for training large models in federated
  learning,'' in \emph{Proceedings of the Thirty-First International Joint
  Conference on Artificial Intelligence, {IJCAI-22}}, L.~D. Raedt, Ed.\hskip
  1em plus 0.5em minus 0.4em\relax International Joint Conferences on
  Artificial Intelligence Organization, 7 2022, pp. 2881--2887, main Track.
  [Online]. Available: \url{https://doi.org/10.24963/ijcai.2022/399}
\BIBentrySTDinterwordspacing

\bibitem{he2020group}
C.~He, M.~Annavaram, and S.~Avestimehr, ``Group knowledge transfer: Federated
  learning of large cnns at the edge,'' \emph{Advances in Neural Information
  Processing Systems}, vol.~33, pp. 14\,068--14\,080, 2020.

\bibitem{cheng2021fedgems}
S.~Cheng, J.~Wu, Y.~Xiao, and Y.~Liu, ``Fedgems: Federated learning of larger
  server models via selective knowledge fusion,'' \emph{arXiv preprint
  arXiv:2110.11027}, 2021.

\bibitem{wu2022exploring}
\BIBentryALTinterwordspacing
Z.~Wu, S.~Sun, Y.~Wang, M.~Liu, Q.~Pan, J.~Zhang, Z.~Li, and Q.~Liu,
  ``Exploring the distributed knowledge congruence in proxy-data-free federated
  distillation,'' \emph{ACM Trans. Intell. Syst. Technol.}, dec 2023, just
  Accepted. [Online]. Available: \url{https://doi.org/10.1145/3639369}
\BIBentrySTDinterwordspacing

\bibitem{anil2018large}
R.~Anil, G.~Pereyra, A.~Passos, R.~Ormandi, G.~E. Dahl, and G.~E. Hinton,
  ``Large scale distributed neural network training through online
  distillation,'' \emph{arXiv preprint arXiv:1804.03235}, 2018.

\bibitem{wang2021knowledge}
L.~Wang and K.-J. Yoon, ``Knowledge distillation and student-teacher learning
  for visual intelligence: A review and new outlooks,'' \emph{IEEE Transactions
  on Pattern Analysis and Machine Intelligence}, 2021.

\bibitem{deng2009imagenet}
J.~Deng, W.~Dong, R.~Socher, L.-J. Li, K.~Li, and L.~Fei-Fei, ``Imagenet: A
  large-scale hierarchical image database,'' in \emph{2009 IEEE conference on
  computer vision and pattern recognition}.\hskip 1em plus 0.5em minus
  0.4em\relax Ieee, 2009, pp. 248--255.

\bibitem{he2020fedml}
C.~He, S.~Li, J.~So, X.~Zeng, M.~Zhang, H.~Wang, X.~Wang, P.~Vepakomma,
  A.~Singh, H.~Qiu \emph{et~al.}, ``Fedml: A research library and benchmark for
  federated machine learning,'' \emph{arXiv preprint arXiv:2007.13518}, 2020.

\bibitem{cifar10}
A.~Krizhevsky, G.~Hinton \emph{et~al.}, ``Learning multiple layers of features
  from tiny images,'' 2009.

\bibitem{he2016deep}
K.~He, X.~Zhang, S.~Ren, and J.~Sun, ``Deep residual learning for image
  recognition,'' in \emph{Proceedings of the IEEE conference on computer vision
  and pattern recognition}, 2016, pp. 770--778.

\bibitem{demlearn-code}
\BIBentryALTinterwordspacing
M.~N. Nguyen, S.~R. Pandey, T.~N. Dang, E.-N. Huh, N.~H. Tran, W.~Saad, and
  C.~S. Hong. [Online]. Available: \url{https://github.com/nhatminh/Dem-AI}
\BIBentrySTDinterwordspacing

\end{thebibliography}
% Generated by IEEEtran.bst, version: 1.14 (2015/08/26)

\vspace{12pt}

\end{document}